\documentclass{aa} 

\usepackage{txfonts}

\begin{document}

\title{Applying the theory of general relativity to reducing
  geodetic VLBI data} 

\author{Titov O.\inst{1}
  \and  Girdiuk A.\inst{2}}

\institute{Geoscience Australia, PO Box 378, Canberra, ACT 2601, Australia\\ \email{oleg.titov@ga.gov.au}
\and Institute of Applied Astronomy, RAS, Kutuzov Quay 10, 191187,
Saint-Petersburg, Russia\\ \email{girduik@ipa.nw.ru} } 


\abstract
{
  We present an alternate formula for calculating
  gravitational time delay.}
  { We use this formula to reduce geodetic
  Very Long Baseline Interferometry (VLBI) data, taking into account
  gravitational effects within the solar system, and to test general
  relativity.
  }
  { The alternate formula was obtained by expanding the
  conventional formula in a Taylor series. We show that the
  gravitational delay can be split into several terms including a term due to the coordinate transformation 
  and terms that are explicitly linked to the light
  deflection angle. }
  { Our formula is compared numerically with the
  conventional formula, and difference in arrival times within 1 ps are found at
  1$^\circ$ from the Sun for a full range of baseline lengths. 
  }
  {We conclude that the standard reduction of geodetic VLBI data for 
  the effects of general relativity is equivalent to  displacing the
  reference radio sources from their original catalogue positions in
  accordance with the classical light deflection formula across the
  whole sky.}


\maketitle

\section{Introduction}
\label{intro}
With geodetic Very Long Baseline Interferometry (VLBI) precise
group delays - the difference in arrival times of radio waves at two
radio telescopes \citep{2012JGeo...61...68S} - can be measured, and very
accurate radio source positions can be obtained ($\sim$ 40
microarcsec \citep{2009ITN....35....1M}). There are numerous astrometric and
geodetic effects that must be accounted for when reducing
high-precision VLBI data. One of these effects is gravitational time
delay, which is caused by the gravitational fields of massive objects
along the observer's line-of-sight \citep{1964PhRvL..13..789S,1967Sci...157..806S}
and is responsible for the measured group delays. These massive
objects also bend the path of light from distant objects, causing an
apparent shift of their position in the sky, as predicted by general
relativity \citep{1916AnP...354..769E}. Within the solar system both the Sun and
Jupiter are massive enough to produce these effects. The
conventional equation for calculating gravitational delay is formulated
in terms of the positions of the radio telescopes within the
barycentric reference frame of the solar system \citep{1990SvA....34....5K,1991resy.coll..256E,1991AJ....101.2306S,1991gvmg.conf..188K}, rather than the
baseline length between the radio telescopes. 
Moreover, the total VLBI delay formula
includes a coordinate term to transform from the geocentric to the
barycentric reference frame.

We propose an alternate gravitational delay formula using a Taylor
series expansion. In Sect. \ref{sec:1} we show how the conventional formula
can be split into a sum of several terms. All the terms are
expressed as a ratio of the baseline length to the Earth barycentric distance.
One of the terms links the gravitational delay and the well-known formula
for the light deflection angle at an arbitrary elongation from
the Sun. 
Another term cancels the coordinate term explicitly presented at the equation of the total VLBI delay. It is more convenient to formulate the
effect of general relativity so that it is free from this complicating coordinate term.

In Sect. \ref{sec:3} a numerical comparison between the new and conventional formulae is presented 
for the cases of the gravitational field of the Sun and Jupiter. \cite{1998AJ....115..361K} compared the astrometric difference in general relativity effects between
the group VLBI delay and optical observations, where the observables are measured in angular quantities. 
It was claimed that the difference is less than 1 microarcsec using some numerical algorithms.
Our analytical approach shows that the two approaches are absolutely equivalent for all elongation angles
(the difference is equal to zero), except for the small area around the gravitational body, which is a special case.

In Sect. \ref{sec:5} we show that the formulae for the post-post-Newtonian effect can be developed using the same approach as for the standard monopole light deflection.

In Sect. \ref{sec:6} we discuss the application of the new formula
to estimate the proper motion of an extragalactic radio source 
induced by the solar gravitational field.
\section{Developing the conventional formula for gravitational delay}

\label{sec:1}

In addition to the three classical tests of general relativity (GR), a fourth test -- the delay of a signal propagating in the solar gravitational field -- was proposed by \cite{1964PhRvL..13..789S} and is known as the Shapiro delay. 
The positions of astronomical instruments on Earth are referenced to the solar system barycentre, and the VLBI delay is equal to the terrestrial time (TT) coordinate time interval between two events of the signal arrival at the first and second radio telescopes \citep{2010ITN....36....1P}. 
The difference between the two Shapiro delays as measured with two radio telescopes at the barycentric distances $\vec{r_1}$ and $\vec{r_2}$ from a body of mass $M$ gives a gravitational delay, $\tau_{grav}$, which must be taken into account during the standard reduction of the high-precision geodetic VLBI data \citep{2010ITN....36....1P},
\begin{equation}\label{shapiro}
\tau_{grav}=\frac{(\gamma+1)GM}{c^3}\ln \frac{|\vec{r_1}|+(\vec{s}\cdot\vec{r_1})}{|\vec{r_2}|+(\vec{s}\cdot\vec{r_2})},
\end{equation}
where $G$ is the gravitational constant, $\vec{s}$ is the unit vector in the direction of observed radio source, $c$ is the speed of light, and $\gamma$ is the parameter of the parametrised post-Newtonian (PPN) formalism \citep{1993tegp.book.....W}, equal to unity in GR. The classical Shapiro delay formula includes the positions of the remote body and the receiver, but the differential delay (\ref{shapiro}) only depends on the positions of the two receivers in the adopted reference system. Therefore, the formula (\ref{shapiro}) is valid for all objects regardless of their distance unless the curvature of the observed light front cannot be ignored. All extragalatic radio sources meet this condition.\\

We introduce the baseline vector, $\vec{b}$, as the difference between the two barycentre radius-vectors of two antennas $\vec{b}=\vec{r_2}-\vec{r_1}$. Then formula (\ref{shapiro}) may be re-written as follows (we recall that $\gamma$ = 1):
\begin{equation}\label{shapiro_2}
\tau_{grav}=\frac{2GM}{c^3}\ln \frac{(\vec{r_2}^2-2(\vec{r_2}\cdot\vec{b})+\vec{b}^2)^{\frac{1}{2}}+(\vec{r_2}\cdot\vec{s})-(\vec{b}\cdot\vec{s})}{|\vec{r_2}|+(\vec{s}\cdot\vec{r_2})}
.\end{equation}
For typical baselines, $|\vec{b}|<<|\vec{r_2}|,$ and we can use the Taylor series expansion $(1+x)^{\frac{1}{2}}\approx1+\frac{x}{2}-\frac{x^2}{8}$ for $x<<1$ to modify (\ref{shapiro_2}) as follows:
\begin{equation}\label{sh_new_2}
\tau_{grav}\approx\frac{2GM}{c^{3}} \ln \left[1+\frac{-(\vec{r_{2}}\cdot\vec{b})/r_{2} -(\vec{b}\cdot\vec{s})+\frac{b^2}{2r_2}-\frac{(\vec{r_{2}}\cdot\vec{b})^2}{2r^3_2}}{r_{2}+(\vec{r_{2}}\cdot\vec{s})}\right]
.\end{equation}
Formula (\ref{sh_new_2}) may be re-written using the approximation $\ln(1+y)\approx y-y^2/2$ for $y<<1$ 
because the baseline length is much smaller than the barycentric distance to the points on Earth
\begin{equation}\label{sh_new_3}
                \begin{array}{lc}
\tau_{grav}\approx \frac{2GM}{c^{3}}&\Bigg[\frac{-(\vec{r_{2}}\cdot\vec{b})/r_{2}-(\vec{b}\cdot\vec{s})+\frac{b^2}{2r_2}-\frac{(\vec{r_{2}}\cdot\vec{b})^2}{2r^3_2}}{r_{2}+(\vec{r_{2}}\cdot\vec{s})} - \\
\\
&-\frac{1}{2}\left(\frac{-(\vec{r_{2}}\cdot\vec{b})/r_{2}-(\vec{b}\cdot\vec{s})}{r_{2}+(\vec{r_{2}}\cdot\vec{s})}\right)^2\Bigg].
                \end{array}
\end{equation}

One has to keep the barycentric coordinates $r_2$ for the second station in the formulae below to provide consistency with the original gravitational delay model (\ref{shapiro}).
\begin{figure}[h!]
\centering
\includegraphics[width=0.5\linewidth]{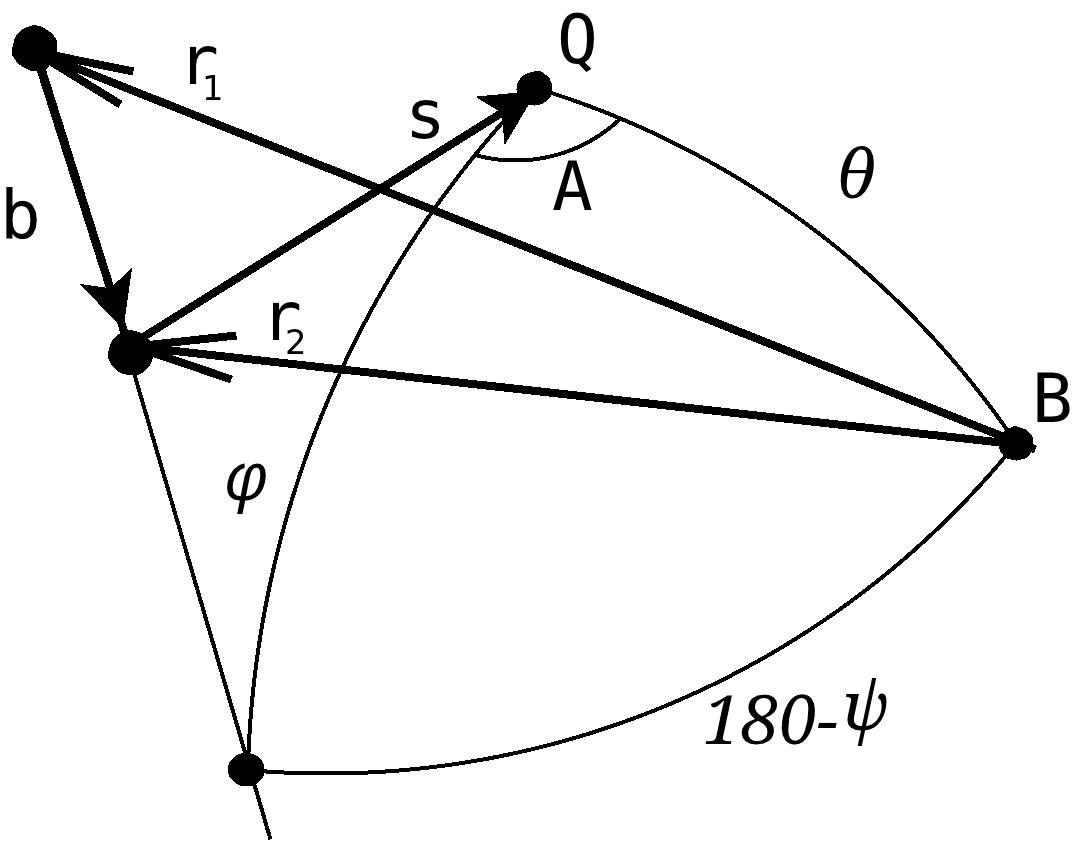}
\caption{Angles $\varphi$, $\psi$, $\theta,$ and $A$, originated from the position of gravitational mass (B), quasar (Q), and baseline vector (\vec{b}). If the Sun plays the role of the gravitational mass, then the point B in Fig. 1 is also the position of the solar system barycentre.}\label{one}
\end{figure}

Converting the three dot products with angles $\varphi,\; \psi, \; \theta$ (Fig.\ref{one}), where 
\begin{equation}\label{angl_mult}
        \begin{array}{lll}
(\vec{b}\cdot\vec{s})&=&|\vec{b}|\cos\varphi,\\
(\vec{b}\cdot\vec{r_2})&=&|\vec{b}||\vec{r_2}|\cos\psi,\\
(\vec{r_2}\cdot\vec{s})&=-&|\vec{r_2}|\cos\theta
        \end{array}
\end{equation}
and changing $|\vec{r_2}|$ to $r_2$  and $|\vec{b}|$ to b for simplicity, one can obtain from Eq. (\ref{sh_new_3})
\begin{equation}\label{vrem}
        \begin{array}{lc}
\tau_{grav}\approx&\frac{2GM}{c^{3}}\Bigg[\frac{-b\cos\psi-b\cos\varphi+\frac{b^2}{2r_2}(1-\cos^2\psi)}{r_2(1-\cos\theta)}-\\
\\
 &-\frac{1}{2}\left(\frac{-b\cos\psi-b\cos\varphi}{r_2(1-\cos\theta)}\right)^2\Bigg].
        \end{array}
\end{equation}
Angle $\theta$ is given here with respect to the second station, that is, $\theta = \theta_2$ for the sake of simplicity.
The difference between $\theta_1$ and $\theta_2$ is discussed in Sect. \ref{sec:3}.
Similar expressions (without terms that include  $\frac{b^2}{r_2}$) have been obtained for geodetic VLBI \citep[e.g.][]{1983Ap&SS..94..233F,1986AJ.....91..650H} and for SIM interferometry mission model delays \citep{2009AstL...35..215T}. 

For the general case of no coplanar vectors (Fig. \ref{one}) 
\begin{equation}\label{psi_angl}
        \begin{array}{c}
\cos\psi=-\cos\varphi\cos\theta-\sin\varphi\sin\theta\cos A,
        \end{array}
\end{equation}
and formula (\ref{vrem}) for arbitrary angles $\varphi$ and $\theta$ is given by
\begin{equation}\label{t_grav0}
                \begin{array}{lcl}
\tau_{grav}&\approx&-\frac{2GMb}{r_2c^{3}}\cos\varphi + \frac{2GMb}{r_2c^{3}}\frac{\sin\varphi\sin\theta\cos A}{1-\cos\theta}+\\
\\
 & &+\frac{GM}{c^{3}}\frac{b^2}{r_2^2(1-\cos\theta)}-\\
\\
 & &-\frac{GM}{c^{3}}\frac{b^2(\cos\varphi\cos\theta+\sin\varphi\sin\theta\cos A)^2}{r_2^2(1-\cos\theta)}-\\ 
\\
 & &-\frac{GM}{c^{3}}\frac{b^2\sin^2\varphi\sin^2\theta\cos^2 A}{r_2^2(1-\cos\theta)^2}-\\
\\
 & &-\frac{GM}{c^{3}}\frac{b^2\cos^2\varphi}{r_2^2}+\frac{GM}{c^{3}}\frac{b^2\sin 2\varphi\sin\theta\cos A}{r_2^2(1-\cos\theta)} .
                \end{array}
        \end{equation}
Using the small-angle approximation ($R<<r_2$),\\
\begin{center}
$\cos\theta\approx1-\frac{R^2}{2r_2^2}$ and $\sin\theta\approx\frac{R}{r_2},$
\end{center}
where $R$ is the impact parameter and is equal to the solar radius for the case of grazing light. 
Therefore
\begin{equation}\label{appr}
                \begin{array}{c}
\frac{b\sin\theta}{r_2(1-\cos\theta)}\approx\frac{2b}{R}, \frac{b^2}{r_2^2(1-\cos\theta)}\approx\frac{2b^2}{R^2}, \frac{b^2\sin\theta}{r_2^2(1-\cos\theta)}\approx\frac{2b^2}{r_2R},
                \end{array}
\end{equation}
and with a $'$typical$'$ baseline of  $\sim$ 6,000 km the first term $\sim$ 160  nsec, the second term is $\sim$ 1.2 ns, and the third term  is $\sim$ 3 ps for grazing light. As we show below, the third term is less than 1 ps for observations at $\theta>1^\circ$ from the Sun. This term and other small terms in (\ref{t_grav0}) may be neglected because VLBI observations are not undertaken closer than 4$^\circ$ from the Sun. Then, the gravitational delay is given by
\begin{equation}\label{t_grav_1}
                \begin{array}{lcl}
\tau_{grav}&\approx&-\frac{2GMb}{r_2c^{3}}\cos\varphi+\frac{2GMb}{r_2c^{3}}\frac{\sin\varphi\sin\theta\cos A}{1-\cos\theta}+\\
\\
 & &+\frac{GM}{c^{3}}\frac{b^2}{r_2^2(1-\cos\theta)}-\frac{GM}{c^{3}}\frac{b^2\cos^2\varphi\cos^2\theta}{r_2^2(1-\cos\theta)}-\\ 
\\
 & &-\frac{GM}{c^{3}}\frac{b^2\sin^2\varphi\sin^2\theta\cos^2 A}{r_2^2(1-\cos\theta)^2},
                \end{array}
\end{equation}
or, finally,
\begin{equation}\label{t_grav_2}
                \begin{array}{lcl}
\tau_{grav}&\approx&-\frac{2GMb}{r_2c^{3}}\cos\varphi+\frac{2GMb}{r_2c^{3}}\frac{\sin\varphi\sin\theta\cos A}{1-\cos\theta}+\\
\\
 & &+\frac{GM}{c^{3}}\frac{b^2(1-\cos^2\varphi\cos^2\theta)}{r_2^2(1-\cos\theta)}-\\
\\
 & &-\frac{GM}{c^{3}}\frac{b^2\sin^2\varphi\sin^2\theta\cos^2 A}{r_2^2(1-\cos\theta)^2}.
                \end{array}
\end{equation}

This equation is sufficient for a picosecond level of accuracy. The first term of Eq. (\ref{t_grav_2}) can be excluded because it cancels out a similar term from the total geocentric delay \citep{2010ITN....36....1P} Eq (11.9). Indeed, this formula (11.9) from the IERS Conventions includes the term

\begin{equation}\label{t_geom}
\tau_{coord} = \frac{(\gamma+1)GM}{c^2r}\frac{(\vec{b}\cdot\vec{s})}{c}
.\end{equation}

Recalling that $\gamma$=1, the resulting effect of general relativity, $\tau_{GR}$,
in the total VLBI delay model is therefore given by the sum of the gravitational delay $\tau_{grav}$ and the coordinate term $\tau_{coord}$

\begin{equation}\label{t_total}
                \begin{array}{lclc}
\tau_{GR} & = & \tau_{grav} + \tau_{coord} \; =  \; \tau_{grav} + \frac{2GM}{c^2r}\frac{(\vec{b}\cdot\vec{s})}{c} &=\\
\\
 &  = &\tau_{grav} + \frac{2GMb}{c^3r}\cos\varphi .&
                \end{array}
\end{equation}

The gravitational delay (\ref{t_grav_2}) includes the barycentric distance of the second station $r_2$ in the denominator, whereas the term (\ref{t_geom}) includes the barycentric distance of the geocentre $r$.
For the case of the Sun, the coordinate term may be as large as 400 ps (see Sect. \ref{sec:2}) and should be considered carefully, but for Jupiter it does not exceed 1 ps.\\

The difference between two terms $\frac{2GM}{c^2r}\frac{(\vec{b}\cdot\vec{s})}{c}$ (\ref{t_geom})  and
$\frac{2GM}{c^2r_{2}}\frac{(\vec{b}\cdot\vec{s})}{c}$ (\ref{t_grav_2}) is less than 0.1 ps, and it may be ignored for the current level of accuracy. Nonetheless, the angles $\theta$ and $\psi$ are to be referred to the vector $r_2$ for correct calculation.

In the original form the gravitational delay is expressed in the BCRS system, whereas the baseline length  in (\ref{t_geom}) is given in  the GCRS system. This means that theoretically, one has to convert the baseline length $b$ and $r_{2}$ into Eq. (\ref{t_grav_2}) from BCRS to GCRS before
comparing the coordinate terms from two different parts. However, in practice, the corresponding correction in time delay does not exceed 1 ps for grazing light and a baseline of 6,000 km.

After all reductions, the total contribution of the relativistic effects to the total VLBI delay, $\tau_{GR}$, can be written as follows:
\begin{figure}[h!]
\centering
        \includegraphics[width=1\linewidth]{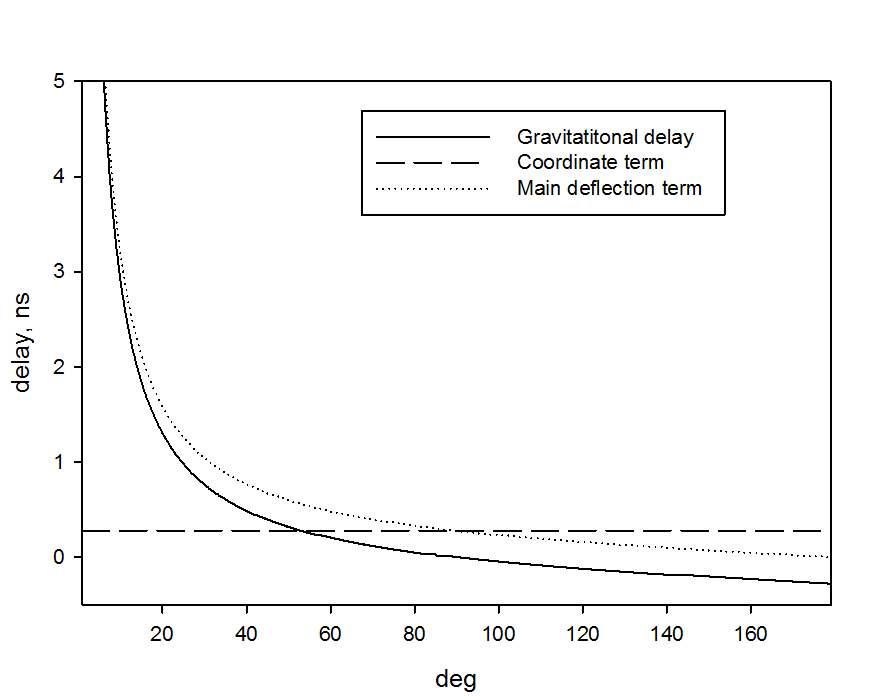}
        \caption{Gravitational delay (\ref{shapiro}), coordinate term, and main deflection term (\ref{t_total}) with respect to the angle $\theta$ (see more details in the text).}
\label{terms}
\end{figure}
\begin{equation}\label{total}
                \begin{array}{lcl}
\tau_{GR}& \approx&\frac{2GMb}{r_2c^{3}}\frac{\sin\varphi\sin\theta\cos A}{1-\cos\theta}+\frac{GM}{c^{3}}\frac{b^2(1-\cos^2\varphi\cos^2\theta)}{r_2^2(1-\cos\theta)}-\\
\\
 & & -\frac{GM}{c^{3}}\frac{b^2\sin^2\varphi\sin^2\theta\cos^2 A}{r_2^2(1-\cos\theta)^2} =: t_1 + t_2 + t_3 .
                \end{array}
\end{equation}

For an arbitrary $\theta$ the gravitational delay (\ref{shapiro}) can be split into several terms in Eq. (\ref{t_total}), and the main deflection term is given by

\begin{equation}\label{arb_teta}
\tau_{GR} = \frac{2GMb}{c^{3}r_2}\frac{\sin\varphi\sin\theta\cos A}{1-\cos\theta} = \alpha_{2}\cdot\frac{b}{c}\sin\varphi\cos A
,\end{equation}

where $\alpha_2$ is the deflection angle at arbitrary $\theta$ \citep{1967Sci...157..806S, 1970ApJ...162..345W},
and is given by

\begin{equation}\label{alpha}
\alpha_{2} =  \frac{2GM}{c^{2}r_2}\frac{\sin\theta}{1-\cos\theta}
.\end{equation}

The minor terms in (\ref{total}) may be ignored for a sufficiently large $\theta$. Therefore (\ref{arb_teta}) may be used as an alternative to the conventional total delay formula for all sources that are far from to a gravitational body. Thus, formulae (\ref{arb_teta}) and (\ref{alpha}) present the deflection angle in a more convenient form for estimation from geodetic VLBI data than previously \citep{1991AJ....102.1879T, 1998AJ....115..361K}. These formulae also prove that the two approaches - based on angular observations and on the VLBI group delay - are equivalent for all ranges of elongation angles, therefore the difference between two types of quantities is equal to zero.
\subsection{Alternative equation for the gravitational fields
of the Sun and Jupiter}

\label{sec:2}

We consider two realistic cases and compare the alternative formula $\tau_{GR}$ (\ref{total}) with the conventional 
formula $\tau_{grav}$ (\ref{shapiro}). There are two bodies in the solar system that are able to cause a significant impact on the observations -
the Sun and Jupiter. The Sun is the more massive of the two bodies, but observations of radio sources closer than $4^\circ$ are very rare. Jupiter is three orders of magnitude less massive, but observations at its limb (i.e., at an extremely small impact parameter of ~70,000 km) are possible. Other massive planets have a weaker effect \citep{2003AJ....125.1580K}.\\

The alternative formula (\ref{total}) consists of three terms: the first term, $t_1$, provides the main contribution to the GR effects, while the two other terms, $t_2$ and $t_3$, contribute much less than $t_1$. However, we show that for very small angles ($\theta$ less than 30$''$) all three terms have a similar contribution.
{At $\theta=180^\circ$ the relativistic contribution, $\tau_{GR}$, to the total delay is equal to zero, whereas the gravitational delay, $\tau_{grav}$,  is up to 400 ps at a $'$standard$'$ baseline of $\sim$ 6,000 km as a result of the coordinate term (\ref{t_geom}) $-\frac{2GMb}{c^3r_2}\cos\varphi$.}\\

{We illustrate this in Fig. \ref{terms}, where the coordinate term (\ref{t_geom}) and the main deflection term (first term in (\ref{total}))} are presented as a function of the impact parameter for $A=180^\circ$ $(\cos A = 1)$ and $\varphi$ = 45$^\circ$ ($\cos\varphi = \sin\varphi=\frac{\sqrt{2}}{2}$), b = 6,000 km, and r = 1 a.u. The coordinate term is always equal to 280 ps because it does not depend on the variable impact parameter $\theta$. The gravitational delay model (\ref{shapiro}) does not contain the angle $\theta$ explicitly, the solid curve in Fig. \ref{terms} is calculated numerically. It is equal to zero at $\theta$ = 90$^\circ$ and -280 ps at $\theta$ = 180$^\circ$. The main deflection term, $t_1$ in (\ref{total}) , is equal to zero at $\theta$ = 180, as was previously noted for the relativistic contribution, $\tau_{GR}$.\\

{To investigate the accuracy of the alternative formula (\ref{total}), one has to sum the original gravitational delay formula (\ref{shapiro}) and the term (\ref{t_geom})}   

\begin{equation}\label{shapiro + geom}
\tau_{GR_2}=\frac{2GM}{c^3}\ln \frac{|\vec{r_1}|+(\vec{s}\cdot\vec{r_1})}{|\vec{r_2}|+(\vec{s}\cdot\vec{r_2})}
+ \frac{2GMb}{rc^{3}}\cos\varphi
.\end{equation}
\begin{figure}[h!]
\centering
        \includegraphics[width=1\linewidth]{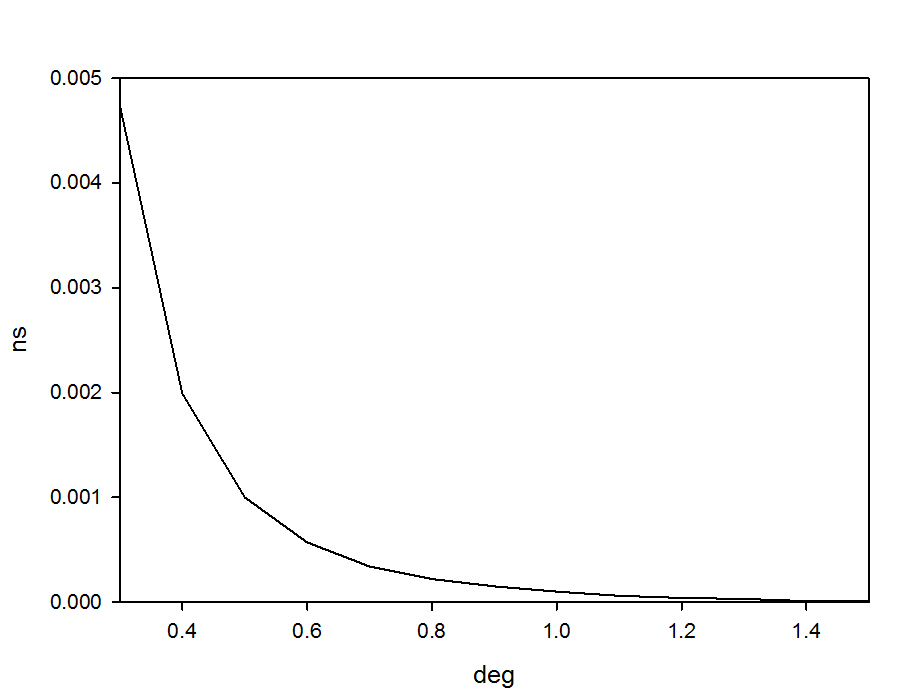}
        \caption{Difference between formulae (\ref{total}) and (\ref{shapiro + geom}) for a baseline of 10,000 km length.}\label{diff}
\end{figure}
Figure \ref{diff} shows the difference between the delays $\tau_{GR}$ (\ref{total}) and $\tau_{GR_2}$ (\ref{shapiro + geom}) for the Sun and a baseline of 10,000 km: the difference between the two formulae only exceeds 1 ps at $\theta$ = 0$^\circ$.5. 
Thus, the difference between the two formulae is negligible for the light propagation in the solar system as measured by geodetic VLBI at the current level of precision.

\begin{figure}[h!]
\centering
        \includegraphics[width=1\linewidth]{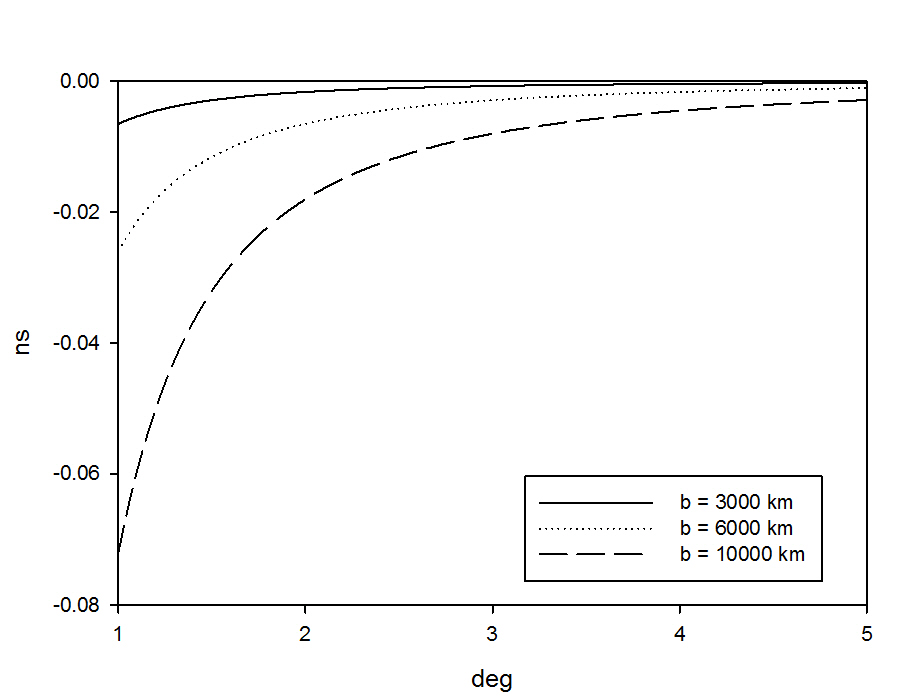}
        \caption{Contribution of the minor terms to the total GR delay for different baseline lengths.}
      \label{minor}
\end{figure}

While the two minor terms, $t_2$ and $t_3$, contribute less than $t_1$, they cannot be ignored completely. The contribution of these terms grows rapidly with baseline length ($\tau \sim b^2$) and is 72 ps at $\theta$ = 1$^\circ$ for $b$ = 10,000 km (Fig \ref{minor}).

\begin{figure}[h!]
                        \begin{minipage}[h]{0.49\linewidth}
                                \centering
                                \includegraphics[width=0.9\linewidth]{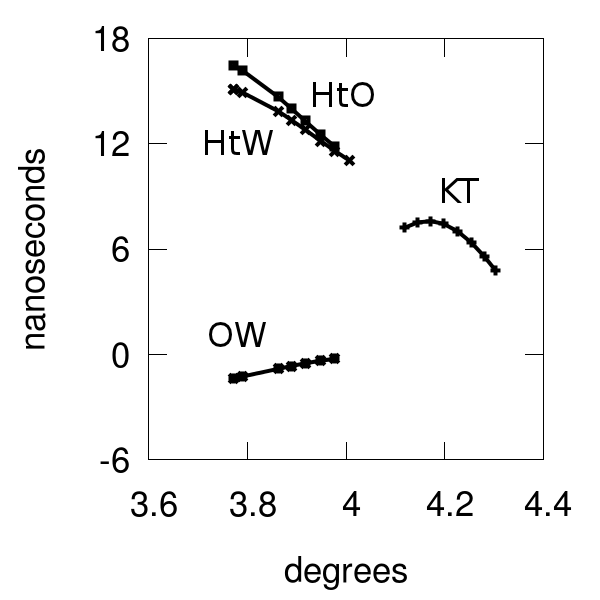}
                        \end{minipage}
                        \hfill
                        \begin{minipage}[h]{0.49\linewidth}
                                \centering
                                \includegraphics[width=1\linewidth]{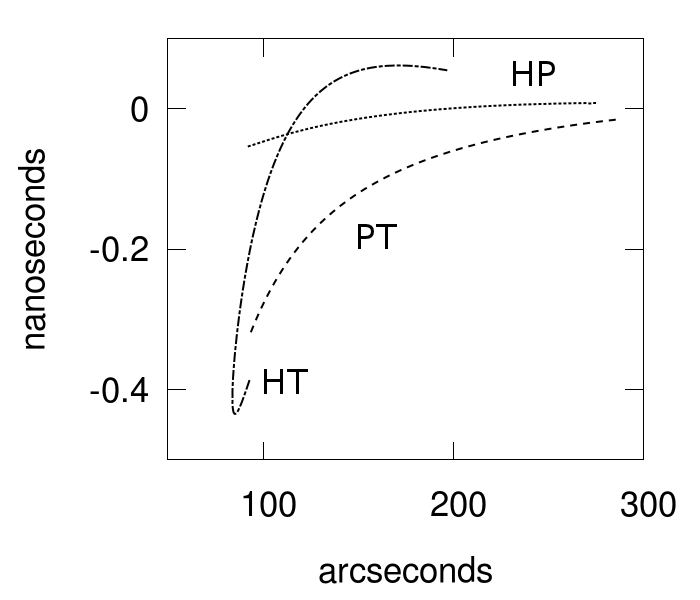}
                        \end{minipage}
                \caption{Main term $t_1$ for  1243-072 (left) and for  1922-224 (right)
                vs $\theta$.}
           \label{ft1t2t3}
                \end{figure}

Herewith, we consider two observational VLBI sessions to separately
investigate the effect of $t_1$ separately during a standard 24-hour session for the Sun and Jupiter. The VLBI experiment R\&D1208 was carried out on 2-3 October, 2012 to track several reference radio sources in close vicinity to the Sun. While the Sun travelled along the ecliptic during the 24-hour session, quasar 1243-072 was tracked in the range of angular distances from of $3^\circ.7$ to $4^\circ.3$. The four curves in the left-hand of plot of Fig. \ref{ft1t2t3} reflect the variation in $t_1$ for four selected VLBI baselines of differing length: Kokee -- Tsukub32 (KT; 5,755 km), \text{HartRAO} -- Wettzell (HtW; 7,832 km), Onsala60 -- Wettzell (OW; 920 km), and HartRAO -- Onsala60 (HtO; 8,525 km) with respect to the angle $\theta$ of the approach of the Sun to the radio source 1243-072. This varies steadily due to the relatively slow apparent motion of the Sun.

The largest solar system planet Jupiter approached the radio source 1922-224 on 18-19 November, 2008.
This event was observed during the 24-hour experiment OHIG60. Four stations - Hobart26, Kokee, Tsukub32, and Parkes  participated in tracking the radio source over only 12 hours because of observing constraints.
The angle $\theta$ between the radio source and Jupiter changed from 1$'$.4 to about 5$'$ within 12 hours. The right plot of Fig. \ref{ft1t2t3} shows the variations in $t_1$ as a function of $\theta$ during this event. This term reaches its maximum of about 400 ps for the longest available baseline (Hobart26 -- Tsukub32 (HT) of $b$ =8,088 km) and becomes negligible after a short period of time. For the shorter baselines Parkes -- Tsukub32 (PT; 7,233 km) and Hobart26 -- Parkes (HP; 1,089 km)), this term is lower in magnitude.

\begin{figure}[h!]
        \begin{minipage}[h]{0.49\linewidth}
        \centering
                \includegraphics[width=0.9\linewidth]{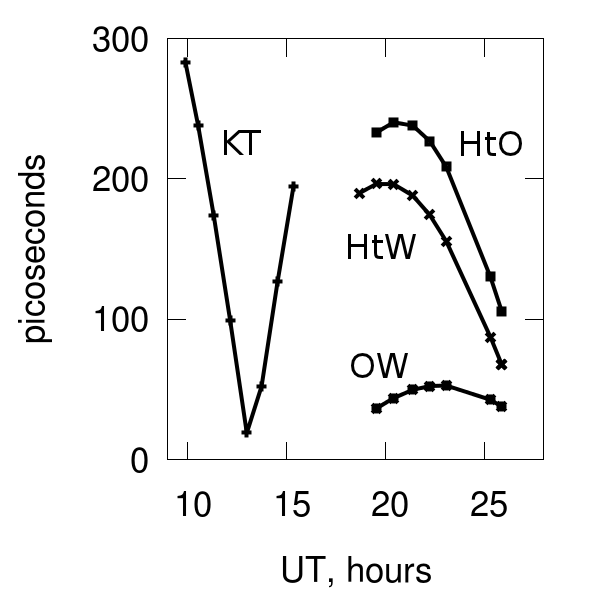}
        \end{minipage}
        \hfill
        \begin{minipage}[h]{0.49\linewidth}
        \centering
                \includegraphics[width=1\linewidth]{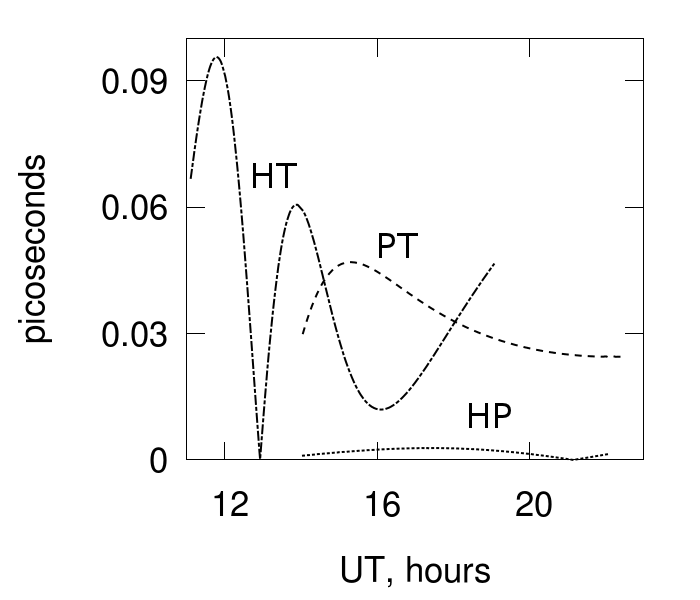}
        \end{minipage}
        \caption{Variations of the coordinate term for 1243-072 (left) and for  1922-224 (right).} 
\label{f2}
\end{figure}

The coordinate term calculated for both events (Fig. \ref{f2}) is significant for the Sun (reaching 300 ps) but is negligible for Jupiter (less than 0.1 ps). In accordance with Eq. (\ref{shapiro + geom}), the coordinate term does not depend on the impact parameter and, therefore, it does not grow at small angles, as seen in Fig. \ref{terms}. All variations in Fig. \ref{f2} are caused by the variations in angle $\varphi$ and baseline length. 

        \begin{figure}
        \begin{center}
        \begin{minipage}[h]{0.49\linewidth}
                \centering
                \includegraphics[width=0.9\linewidth]{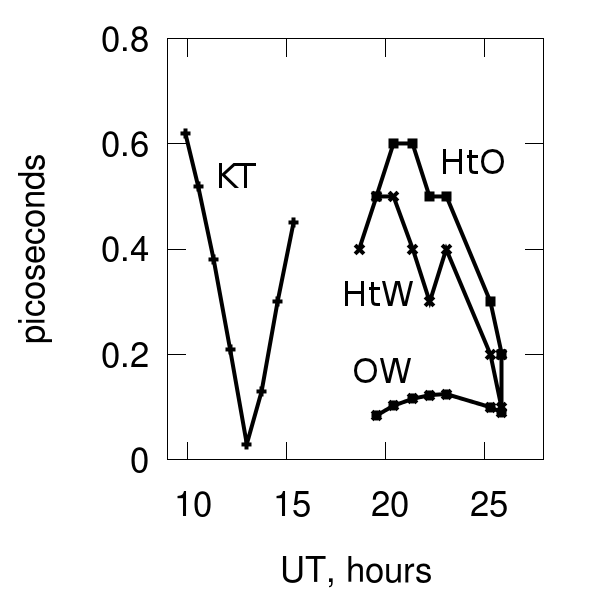}
        \end{minipage}
        \hfill
        \begin{minipage}[h]{0.49\linewidth}
                \centering
                \includegraphics[width=1\linewidth]{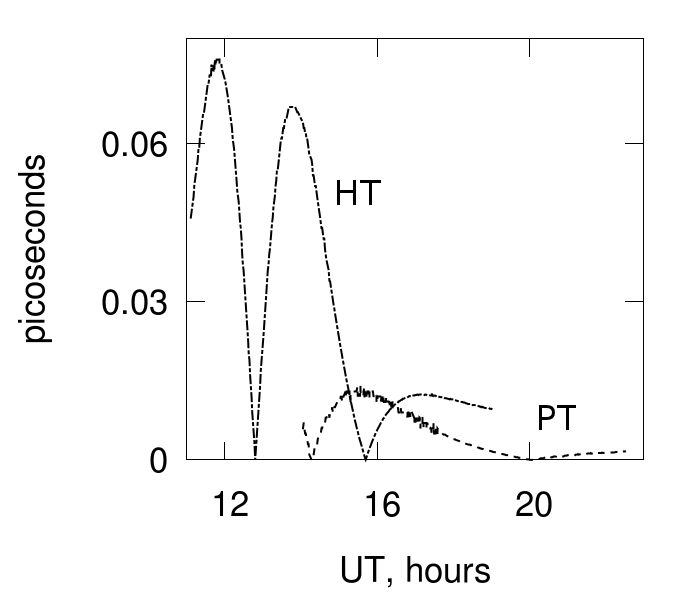}
        \end{minipage}
        \caption{Comparison of the two models for the approach of the Sun to 1243-072 (left) and of Jupiter to 1922-224 (right). The difference for baseline Hobart26 -- Parkes in the right plot is not distinctive for this scale.}\label{j2}
        \end{center}
        \end{figure}

Figure \ref{j2} illustrates the residuals between Eqs. (\ref{total}) and (\ref{shapiro + geom})  for these two events. The difference does not exceed 1 ps for either case. This proves that Eq. (\ref{total}), proposed in this paper, can be used to model the relativistic effects along with the conventional model (\ref{shapiro + geom}).
\section{Small-angle approximation.}
\label{sec:3}
The monopole effect of the general relativity results in a circular motion of the deflected object \citep{2001MNRAS.323..952S, 2007PhRvD..75f2002K}. Therefore, the radio sources draw a regular pattern on the sky due to the light deflection. Here we consider an impact of the minor terms in Eq. (\ref{total}) on the formula for the light deflection, in particular on the intraday time-scale, if a large planet such as Jupiter passes by a reference radio sources very quickly. In the small-angle approximation, Eq. (\ref{appr}), the effect of GR (\ref{total}) is given by
\begin{equation}\label{Einstein}
                \begin{array}{lcl}
\tau_{GR} & = &\frac{4GM}{c^{2}R_{2}}\frac{b}{c}\sin\varphi(\cos A - \frac{b}{2R_{2}}\sin\varphi\cos 2A) =\\
\\
& = & \alpha'_{2} \cdot \frac{b}{c}\sin\varphi(\cos A - \frac{b}{2R_{2}}\sin\varphi\cos 2A),
                \end{array}
\end{equation}
where $\alpha'_{2}$ is the deflection angle for light propagated through a gravitation field \citep{1916AnP...354..769E} for the second station 
\begin{equation}\label{Einstein_classic}
\alpha'_{2} =  \frac{4GM}{c^{2}R_{2}}
\end{equation}
and $R_{2} = r_{2}\cos\theta_{2}$.
\begin{table}
\caption{Einstein deflection angle and the secondary deflection angle for the Sun.}
\label{tab:1}
\centering
\begin{tabular}{rll|cc}
\hline\hline
$R_2$, km & $\theta$ & $\alpha'_2$ & \multicolumn{2}{c}{$\alpha''_2$, mas}  \\
      &          &   & $b$ = 3,000 km & $b$ = 10,000 km \\
\hline
700,000    & 0$^\circ$.25 & 1$''$.75 & 3.75 & 12.5 \\
2,800,000  & 1$^\circ$.0  & 0$''$.44 & 0.23 & 0.80 \\
5,600,000  & 2$^\circ$.0  & 0$''$.22 & 0.06 & 0.20 \\
14,000,000 & 5$^\circ$.0  & 0$''$.09 & 0.01 & 0.03 \\
\hline
\end{tabular}
\end{table}
\begin{table}
\caption{Einstein deflection angle and the secondary deflection angle for Jupiter.}
\label{tab:2}       
\centering
\begin{tabular}{rrccc}
\hline\hline
$R_2$, km & $\theta$ & $\alpha'_2$, mas & \multicolumn{2}{c}{$\alpha''_2$, mas}  \\
          &          &                  & $b$ = 3,000 km & $b$ = 10,000 km \\
\hline
 70,000 & 20$''$ & 16 & 0.34 & 1.14 \\
210,000 & 1 $'$  &  5 & 0.04 & 0.13 \\
420,000 & 2 $'$  &  2 & 0.01 & 0.03 \\
\hline
\end{tabular}
\end{table}
                \begin{figure}[h!]
                \hspace*{-0.2cm}
                \begin{minipage}[h]{0.49\linewidth}
                        \centering
                                \includegraphics[width=1.1\linewidth]{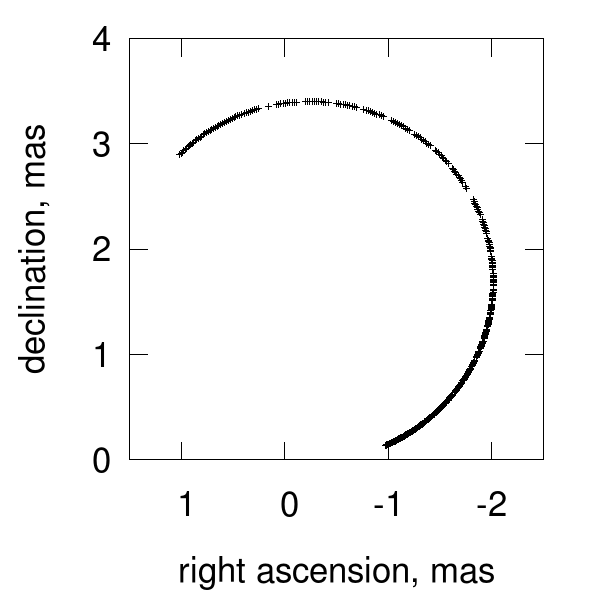}
                        \end{minipage}
                        \hfill
                        \begin{minipage}[h]{0.49\linewidth}
                                \centering
                                \includegraphics[width=1.1\linewidth]{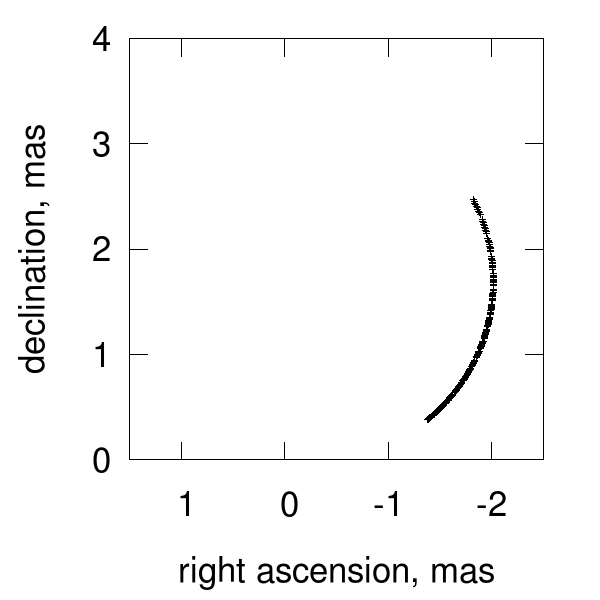}
                        \end{minipage}
                        \vfill
                \hspace*{-0.2cm}
                        \begin{minipage}[h]{0.49\linewidth}
                                \centering
                                \includegraphics[width=1.1\linewidth]{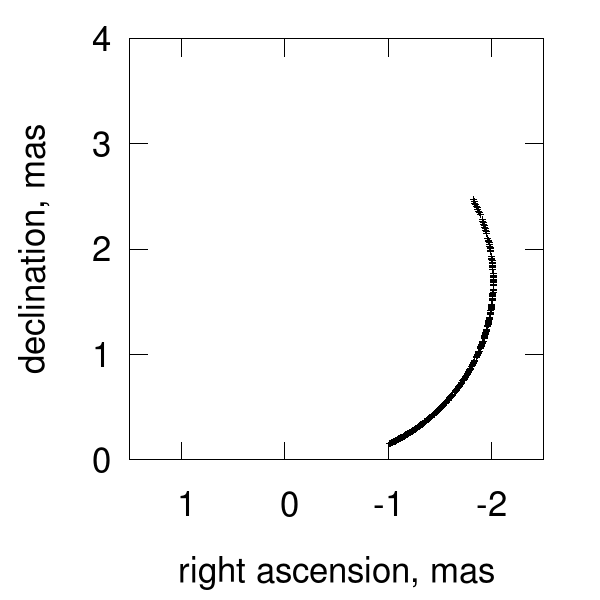}
                        \end{minipage}
                        \hfill
                        \begin{minipage}[h]{0.49\linewidth}
                                \centering
                                \includegraphics[width=1.1\linewidth]{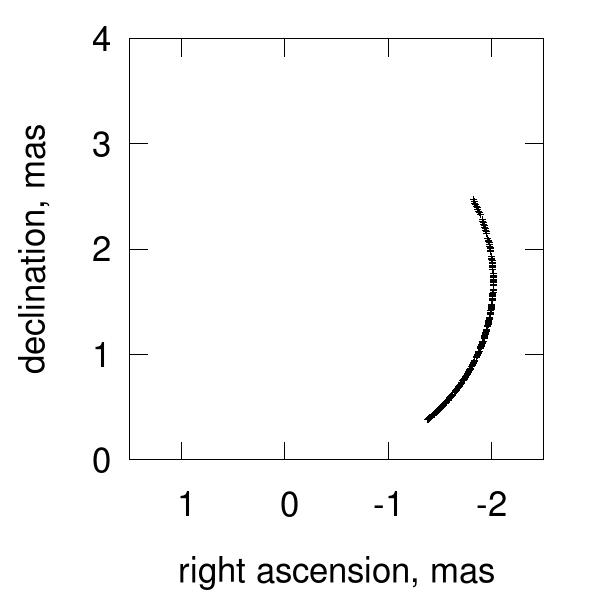}
                        \end{minipage}
                        \caption{Arc for baselines Hobart26-Tsukub32 (upper left), Kokee-Tsukub32 (upper right),  Hobart26-Parkes (bottom left), Kokee-Parkes (bottom right).}
           \label{f8}
        \end{figure}
Re-arranging formula (\ref{Einstein}) provides an expression for the deflection angle, $\alpha'_{2}$, for an arbitrary baseline length in the small-angle approximation as follows:
\begin{equation}
                \begin{array}{lclc}
\alpha = \frac{c\tau_{GR}}{b\sin\varphi\cos A} & = &\frac{4GM}{c^{2}R_{2}}(1 - \frac{b}{2R_{2}}\frac{\cos 2A \sin \varphi}{\cos A}) &= \\ 
\\
& = &\alpha'_{2} (1 - \frac{b}{2R_{2}}\frac{\cos 2A \sin \varphi}{\cos A})&=\\
\\
& = &\alpha'_{2} + \alpha''_{2}.&
                \end{array}
\end{equation}

Thus, once in single-telescope mode, only the main angle (\ref{Einstein_classic}) is presented; in the two-telescope mode, the secondary deflection angle 
\begin{equation}
\alpha''_{2} = - \frac{2GM}{c^{2}R_{2}} \frac{b}{R_{2}} \frac{\cos 2A \sin \varphi}{\cos A}
\end{equation}
represents the difference between angles $\alpha'_{2}$ and $\alpha'_{1}$, if $\alpha'_{1} =  \frac{4GM}{c^{2}R_{1}}$
and $R_{1} = r_{1}\cos\theta_{1}$. Even though the parallactic shift of the Sun 
$\theta_{1} - \theta_{2}$ is about 8$''$ for $b = $ 6,000 km, this difference between $\alpha'_{1}$ and $\alpha'_{2}$, is not negligible in the small-angle approximation. The formula shows that the secondary deflection angle depends on the baseline length, but it reduces to Einstein's classical expression (\ref{Einstein_classic}) for a zero-length baseline (i.e. $b = $0).

Table \ref{tab:1} shows the meaning of the angle $\alpha''_{2}$ for the Sun and Table \ref{tab:2} for Jupiter with different combinations of the parameters $\theta_{2}$ and $R_{2}$. 
This effect rapidly grows as $b$ is decreasing and may be detected with a long-baseline radio-interferometer at a very close approach of the Sun ($< 5^\circ$) or large planets ($< 1'$) to a distant radio source.

The angle $\alpha'_2$ can change comparatively fast when Jupiter passing by a radio source. An example of the fast variations
of $\alpha'_2$  is shown in  Fig. \ref{f8} for the approach of Jupiter and the radio source 1922-224 discussed in Sect. \ref{sec:2}. The corresponding time delays as shown in the right plot of Fig. 5 are different, but the light deflection angles in Fig 8. for all baselines are the same.

\section{Post-post-Newtonian effect}
\label{sec:5}
As shown by \citep{1991gvmg.conf..188K}, the main term of the post-post-Newtonian effect
is given by
\begin{equation}
                \begin{array}{lcl}
\tau_{PPN} &=& \frac{4G^2M^2}{c^5}\left[\frac{4}{|\vec{r_1}|+(\vec{r_1}\cdot \vec{s})}-\frac{4}{|\vec{r_2}|+(\vec{r_2}\cdot \vec{s})}\right] 
.\end{array}
\end{equation}
Using the mathematical transformation of Eqs. (\ref{shapiro_2}) and (\ref{angl_mult}),
\begin{equation}\label{ppn2}
                \begin{array}{lcl}
\tau_{PPN} &=& \frac{4G^2M^2}{c^5}\left[\frac{4}{r_2(1-\cos\theta)\left(1-\frac{b\cos\psi+b\cos\varphi}{r_2(1-\cos\theta)}\right)}-\frac{4}{r_2(1-\cos\theta)}\right]
.\end{array}
\end{equation}
If the baseline length is negligible with respect to the barycentric distance ($b<<r_2$), one could recall $x=\frac{b\cos\psi+b\cos\varphi}{r_2(1-\cos\theta)}$ and apply the approximation $\frac{1}{1-x} = (1+x)$ using Eq. (\ref{psi_angl}) to solve the spherical triangle (Fig. 1) and to develop (\ref{ppn2}) as follows:
\begin{equation}\label{ppn}
                \begin{array}{lclc}
\tau_{PPN} & = & \frac{4G^2M^2b}{c^5}\frac{-cos\varphi\cos\theta-\sin\varphi\sin\theta\cos A + \cos\varphi}{r_2^2(1-\cos\theta)^2} &= \\
\\
 & = & \frac{4G^2M^2b}{c^5}\frac{\cos\varphi}{r_2^2(1-\cos\theta)}-\frac{4G^2M^2b}{c^5}\frac{\sin\varphi\sin\theta\cos A}{r_2^2(1-\cos\theta)^2}.&
                \end{array}
\end{equation}
In the small-angle approximation ($R<<r_2$) (\ref{ppn}) is given by
\begin{equation}\label{ppn_small_angle}
                \begin{array}{lcl}
\tau_{PPN} &=& \frac{4G^2M^2b}{c^5}\frac{\cos\varphi}{R^2} -  \frac{16G^2M^2b}{c^5}\frac{r_2\sin\varphi\cos A}{R^3}
.\end{array}
\end{equation}
For the grazing light the first term in Eq. (\ref{ppn_small_angle}) is about 1 ps, whereas the second one can reach 300 ps at a 'standard' baseline of 6,000 km \citep{1991gvmg.conf..188K}. 

When the first term in Eq. (\ref{ppn}) is negligible, we can introduce a new variable
\begin{equation}\label{ppn_angl}
                \begin{array}{lcl}
\alpha_{PPN} &=& \frac{4G^2M^2}{c^4}\frac{\sin\theta}{r_2^2(1-\cos\theta)^2}
				\end{array}
\end{equation}
and re-write the second term of Eq. (\ref{ppn}) as follows:
\begin{equation}\label{ppn_tau}
                \begin{array}{lcl}
\tau_{PPN} &=& - \alpha_{PPN}\frac{b}{c}\sin\varphi\cos A
.				\end{array}
\end{equation}
It is obvious that the variable $\alpha_{PPN}$ plays the role of the post-post-Newtonian deflection angle at arbitrary $\theta$, and converting between the deflection angle and the corresponding part of gravitational delay uses the same factor $\frac{b}{c}\sin\varphi\cos A$  for the post-Newtonian effect as in Eqs. 
(\ref{arb_teta}) and (\ref{alpha}). It enables us to present the sum of the two delays $\tau_{GR}$ (\ref{arb_teta}) and $\tau_{PPN}$ (\ref{ppn_tau}) in the form of two angles, Eqs. (\ref{Einstein_classic}) and (\ref{ppn_angl}), multiplied by $\frac{b}{c}\sin\varphi\cos A$, that is,
\begin{equation}\label{all}
\tau_{GR}+\tau_{PPN} = (\alpha_{GR}+\alpha_{PPN})\frac{b}{c}\sin\varphi\cos A
.\end{equation}

Finally, we recall that the second term of Eq. (\ref{ppn}) corresponds to formula (11.14) from the IERS Conventions \citep{2010ITN....36....1P}.

\section{Proper motion of extragalactic radio sources induced by the Sun}
\label{sec:6}
The solar gravitational field will result in a proper motion of extragalactic radio sources if the effects of general relativity are not accounted for in VLBI data reduction. 
The rapid change of the observed positions during a typical 24-hour VLBI session may be detected for observations within 10$^\circ$ from the Sun. 

The partial derivative $\frac{d\tau_{GR}}{d\theta}$ is given by
\begin{eqnarray}\label{dtau_dtheta}
\frac{d\tau_{GR}}{d\theta} = -\frac{2GM}{c^{3}}\frac{b\sin\varphi\cos A}{r(1-\cos\theta)}
.\end{eqnarray}

For an arbitrary angle $\theta,$ the proper motion $\mu = \frac{d\alpha}{dt}$ (where $\alpha$ is from (\ref{arb_teta})) may be presented by re-arranging Eq. (\ref{dtau_dtheta}) as follows:
\begin{eqnarray}\label{motion}
\begin{split}
\mu  = \frac{d\alpha}{dt} \approx \frac{c}{b\sin\varphi\cos A}\frac{d\tau_{GR}}{dt} & = 
\frac{c}{b\sin\varphi\cos A}\frac{d\tau_{GR}}{d\theta}\frac{d\theta}{dt}=\\
& = - \frac{2GM}{c^{2}r}\frac{1}{(1-\cos\theta)}\frac{d\theta}{dt}
\end{split}
.\end{eqnarray}

\begin{figure}[h!]\label{pmplot}
        \centering
        \includegraphics[width=1\linewidth]{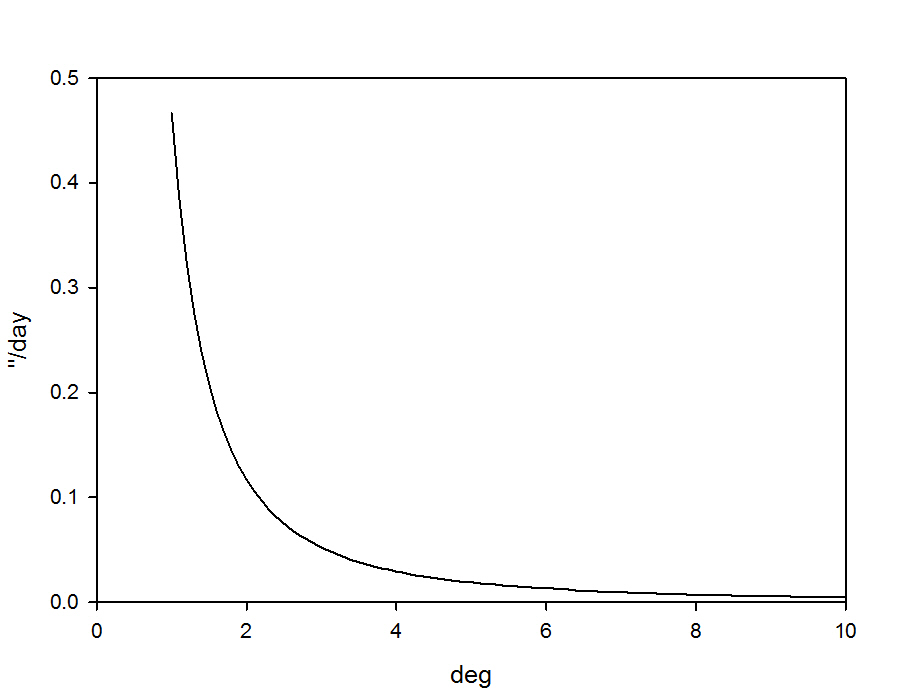}
        \caption{Daily proper motion calculated from Eq. (\ref{motion}) near the Sun.}
     \label{f9}
\end{figure}

Figure \ref{f9} illustrates Eq. (\ref{motion}) for a range of angles $\theta$ from 0$^\circ$ to 10$^\circ$. It is obvious that all distant radio sources across the sky will display a circular motion if general relativity is not accounted for in VLBI data reduction. The magnitude of the effect is proportional to $\frac{\sin\theta}{1 - \cos\theta}$. In particular, a source near the ecliptic poles $\theta = 90^{\circ}$ will draw a circle of 4 mas in size. The daily displacement between the consecutive days will be equal to $\frac{2GM}{c^{2}r}\frac{d\theta}{dt}$. For the Sun, $\frac{d\theta}{dt}=0.0174,$ and the daily displacement is $\approx 0.07$ mas.
For all other sources, the daily displacement will be proportional to the factor $\frac{1}{(1-\cos\theta)}$ in Eq. (\ref{motion}).

In the small-angle approxiation, Eq. (\ref{motion}) may be written as follows:

\begin{eqnarray}\label{motion_2}
\mu  = - \frac{4GMr}{c^{2}R^2}\frac{d\theta}{dt} = - \alpha\frac{r}{R}\frac{d\theta}{dt}
.\end{eqnarray}

For background radio sources that are co-planar with the plane of the body, the apparent motion, $\mu$, rapidly grows from 0.5 $\mu$as/year at $\theta = 10^{\circ}$, to 1.6 $\mu$as/year at $\theta = 5^{\circ}$ and 40 $\mu$as/year at $\theta = 1^{\circ}$. Therefore, a systematic pattern of proper motion in a particular area of the sky may serve as an indicator of a hidden mass in this direction.\\

\section{Conclusion}

The effects of general relativity are explicitly contained in both components of the total VLBI delay model - the gravitational delay and the coordinate term with a factor of $1-\frac{2GM}{c^{2}r}$. While the former component uses the individual barycentre positions of the radio telescopes to calculate the effect, the latter component is expressed in terms of the baseline between the radio telescopes. Coupling between both parts has not been investigated until now.

The Shapiro effect as measured by radars and the deflection of light measured with traditional astronomical instruments are considered two independent tests of general relativity. We showed that the total group delay model joins the two tests within one observational technique - geodetic VLBI. The gravitational delay that traditionally originated from the Shapiro effect is linked to the light deflection angle. Therefore, the two approaches, VLBI delay and angular, are absolutely equivalent.

For almost all realistic situations this angle does not depend on the baseline length, thus, a standard geodetic VLBI interferometer acts as a traditional astronomical instrument. In addition, the coordinate term explicitly presented in the conventional total delay model ceases to exist because it is compensated by the same effect in the gravitational delay with opposite sign. Thus, the proposed alternative version of the general relativity contribution to the total VLBI group delay model is free of the coordinate effects. Therefore, the two approaches, time delay and angular, are absolutely equivalent.

The final equation of the general relativity contribution expressed in angular terms also comprises two smaller terms that are significant at very small angular separation between the deflecting body and distant radio source. These terms may be considered as an increment in the light deflection angle due to the additional time delay for the propagation of the light from station 1 to station 2. This effect becomes significant at $\frac{b}{R}>0.1$.

It is unlikely that the effect will be detected with ground-based interferometers, but it may be detected with space-baseline 
interferometers, for example, RadioAstron with $b$ up to 300,000 km \citep{2013ARep...57..153K}.
   
\begin{acknowledgements}
We thank Craig Harrison for proofreading the manuscript and valuable comments. We are also grateful to Slava Turyshev, Sergei Kopeikin, Sergei Klioner, Laura Stanford, Benedikt Soja, Natalia Pavlovskaya, Nicole Capitaine, Sebastien Lambert, Adrien Bourgoine,  Pierre Teyssandier, Aurelien Hees, Christophe Le Ponchin-Lafitte, Vladimir Zharov and Mihail Sazhin for useful discussion on the mathematical details of general relativity. Oleg Titov was supported by the Russian fund 'Dynasty' for two short-term working visits to Saint-Petersburg University in 2011 and 2013. A special "hand made" observing schedule was prepared by Dirk Behrend (GSFC) to track the radio source 1922-224 on 18-19 November, 2008.\\
This paper is published with the permission of the CEO, Geoscience Australia.
\end{acknowledgements}

\end{document}